\documentclass[twocolumn,showpacs,preprintnumbers,amsmath,amssymb]{revtex4}

\usepackage{graphicx,psfrag}
\usepackage{amssymb}
\usepackage{amsmath}
\usepackage[squaren,Gray,textstyle]{SIunits}
\usepackage{dcolumn}
\usepackage{bm}

\def\nc{\newcommand}

\def\lsim{\mathrel{\raise.3ex\hbox{$<$\kern-.75em\lower1ex\hbox{$\sim$}}}}
\def\gsim{\mathrel{\raise.3ex\hbox{$>$\kern-.75em\lower1ex\hbox{$\sim$}}}}
\nc{\half}{\frac{1}{2}}
\nc{\shalf}{\ensuremath{\textstyle \frac{1}{2}}}

\nc{\deldag}{\mathbin{\partial\mkern-10.5mu\big/}}
\nc{\kdag}{\mathbin{k\mkern-10mu\big/}}
\nc{\Pdag}{\mathbin{P\mkern-10mu\big/}}

\nc{\beq} {\begin{equation}}
\nc{\eeq} {\end{equation}}
\nc{\beqa}{\begin{eqnarray}}
\nc{\eeqa}{\end{eqnarray}}

\def\half{\frac{1}{2}}
\def\l({\left(}
\def\r){\right)}

\begin{document}


\title{On white dwarfs and neutron stars in Palatini f(R) gravity}
\author{Vappu Reijonen${}^{1}$}
   \email{vappu.reijonen@helsinki.fi}

\affiliation{
    ${}^{1}$Helsinki Institute of Physics and Dept.~of Physics,
        P.O.~Box 64, FIN-00014 University of Helsinki, Finland}

\date{\today}

\begin{abstract}

In Palatini $f(R)$ gravity, the parameters of the Schwarzschild - de Sitter solution as well as the whole interior solutions of compact objects are expected to change when compared to general relativity. We solve the Palatini field equations numerically in the case of the models $f(R) = R + \alpha R^2$ and $f(R) = R - \mu^4/R$, and using the equation of state of Fermi gas. We show how the density profiles and the prediction for the maximum masses of white dwarfs (the Chandrasekhar limit) and neutron stars (the Tolman-Oppenheimer-Volkoff limit) are altered, and thereby conclude that observations on compact stars may be used to exclude alternative gravity models.

\end{abstract}

\pacs{98.80.-k, 04.50.+h, 95.35.+d}

\maketitle

%
%

\section{Introduction}
\label{sec:intro}

In general, general relativity is regarded as an effective theory of gravity, its breakdown being practically inevitable at the Planck scale. Whether the breakdown could show at lower scales has been a topical question for long. Classical extensions to Einstein's theory have not only been studied in the context of inflation and early universe, but also to account for dark energy and dark matter~\cite{Nojiri:2003ft}. Moreover, any suggested alternative gravity model should agree with the Solar System constraints and other experimental astrophysical data~\cite{Will:2005va}.

The $f(R)$ gravity models have been among the simplest and most popular extensions of the Einstein-Hilbert action. Originally suggested as models of inflation~\cite{Starobinsky:1980te}, they have gained interest as gravitational dark energy during past few years~\cite{vollick}. The action
\begin{equation}
    S = \frac{1}{16 \pi G} \int {\rm d}^{4}x \sqrt{-g} f(R) + S_{\rm m}
\label{eq:action}
\end{equation}
generalizes the gravitational sector to involve non-linear interactions in the Ricci scalar: setting $f(R) = R - 2\Lambda$ corresponds to the canonical Einstein-Hilbert action with a cosmological constant $\Lambda$.

In general relativity, the metric $g$ is usually taken as the only free gravitational field. The case, where both the metric and the connection $\Gamma$ are dynamical variables, was also studied very early after the formulation of general relativity~\cite{palatini}. The fundamental theorem of Riemannian geometry states that on any (pseudo-)Riemannian manifold there is a unique torsion-free connection for the given metric $g$ \emph{that satisfies metric compatibility}, and that connection is the Levi-Civita (Christoffel) connection. The compatibility condition $\nabla g = 0$ for the only traditionally (physically) allowed covariant derivative equals to demanding that test (physical) particles follow "the shortest paths on the manifold" ~\cite{mtw}. More formally, geodesics (defined by the connection) coincide with the extremals of the length function (defined by the metric) on the manifold. Choosing the covariant derivative with respect to the Levi-Civita connection, the act of parallel transport is intuitive: it preserves the inner products, which includes the norm and the sense of orthogonality.

While this all might sound reasonable, the harsh fact is that there is \emph{no unique way} to compare two tensors if they are elements of different tangent spaces. In fact, it can be said that we do not know if the geodesics truly coincide with the shortest paths on our spacetime manifold. What we do know very accurately, by experiment~\cite{Su:1994gu}, is that all local effects of gravitation disappear, which imposes the condition
\beq
\Gamma^{\rho}_{\mu \nu}(P_0) = 0
\eeq
for the connection coefficients in the local Lorentz frame at any spacetime point $P_0$. In all, there seems to be no immediately clear, physical reason to select the Levi-Civita connection from all the possible connections on the manifold.

The formulation, where both the metric and connection are left as free variables, is referred to as \emph{Palatini formalism}, while insisting on keeping the connection as the Levi-Civita connection is called \emph{metric formalism}. Starting from varying the action, however, the Einstein-Hilbert action (with or without a cosmological constant) yields the connection to be the Levi-Civita connection, \emph{if} the connection is torsion-free. In the case of the general $f(R)$ action~(\ref{eq:action}) this is not the case: the metric and Palatini formalisms give very different gravity out of the same action.

Perhaps somewhat surprisingly, many \emph{metric} $f(R)$ models for dark energy, such as $f(R) = R - \mu^4/R$ with $\mu^4 \sim 10^{-104}$m$^{-4}$, are shown to contradict the Solar System measurements~\cite{erickcek}. These models fail to produce the experimentally bound exterior solution outside a Newtonian star. The corresponding Palatini models, however, reduce to the limit of general relativity at the Solar System and are perfectly consistent with the experiments~\cite{kainulainen}. Nevertheless it seems that the dark energy Palatini $f(R)$ models fail to produce the observed large scale structure in the universe~\cite{tomi}.

Moreover, models with extra terms such as $f(R) = R + \alpha R^2$ are not restricted by weak field effects. Such terms are expected to arise in quantizing gravity. In this work we study whether and how the structure of stars with strong gravitational fields is changed in the framework of $f(R)$ models in comparison with general relativity. While the literature on the microphysics of compact stars is vast (that is on the effects that kick in gradually as the state of matter changes at high energies), the issue of changing effective gravity has been rather neglected.

%
%

\section{f(R) gravity}
\label{sec:palatini}

The field equations of Palatini gravity are attained by varying the action~(\ref{eq:action}) with respect to the metric and the connection,
\beqa
    F(R)R_{\mu \nu} - \half f(R) g_{\mu \nu} = 8 \pi T_{\mu \nu}
\label{eq:field1} \\
    \nabla_{\rho}\left[ \sqrt{-g} F(R) g^{\mu \nu} \right] = 0,
\label{eq:field2}
\eeqa
respectively, where $F \equiv \partial f / \partial R$ (in general relativity, $F = 1$, Eq.~(\ref{eq:field1}) identifies as the Einstein equation and Eq.~(\ref{eq:field2}) as metric compatibility). The covariant derivative $\nabla_{\rho}= \nabla_{\rho}(\Gamma)$ and the Ricci tensor $R_{\mu \nu} = R_{\mu \nu}(\Gamma)$ (and thus the Ricci scalar via $R = g^{\mu \nu} R_{\mu \nu}(\Gamma)$) are functions of the free connection. From the latter equation, however, we see immediately that the connection is the Levi-Civita connection \emph{for the conformally related metric} $h_{\mu \nu} = F g_{\mu \nu}$,
\beqa
& & \Gamma^{\rho}_{\mu \nu} = \big\{ \stackrel{\rho}{\mu \nu} \big\} [h] \\
& = & \big\{ \stackrel{\rho}{\mu \nu} \big\} [g] + \frac{1}{2F}g^{\rho \sigma}(g_{\sigma \mu} \partial_{\nu} F + g_{\sigma \nu} \partial_{\mu} F - g_{\mu \nu} \partial_{\sigma} F). \nonumber
\eeqa
In the metric formalism the connection is taken to be the Levi-Civita connection of the metric $g$. The field equation is obtained by varying the action~(\ref{eq:action}) with respect to the metric only,
\beq
    F R_{\mu \nu} - \frac{1}{2} f g_{\mu \nu}
                 - \nabla_\mu\nabla_\nu F+ g_{\mu \nu}\Box F
               = 8\pi G T_{\mu \nu},
    \label{eq:fieldmetric}
\eeq
where the Ricci tensor and the covariant derivative are now defined directly by the Levi-Civita connection.

Taking the traces of the field equations~(\ref{eq:field1}) and~(\ref{eq:fieldmetric}) highlights the difference between the Palatini and metric gravities. In the Palatini case the trace equation relates the Ricci scalar to the energy momentum tensor \emph{algebraically},
\beq
  FR - 2f = 8 \pi T, \:
\label{eq:trace}
\eeq
which is also the case in general relativity. In the metric formalism the trace equation is a second order differential equation for the Ricci scalar,
\beq
3\Box F + FR - 2f = 8\pi T, \;
\label{eq:tracemetric}
\eeq
that is, the equation~(\ref{eq:tracemetric}) is a \emph{fourth} order differential equation for the metric tensor. This feature allows for several vacuum solutions ($T_{\mu \nu}=0$).

Taking the covariant derivative of the left hand side of equations~(\ref{eq:field1}) and~(\ref{eq:fieldmetric}) yields Bianchi identities, which is shown in detail in~\cite{tomi2}. The continuity equation takes the same form in both metric and Palatini formalisms,
\beq
\nabla_{\mu} T^{\mu \nu} = 0,
\label{eq:cont}
\eeq
where $\nabla_{\mu}$ is the covariant derivative with respect the Levi-Civita connection.

\section{The Palatini source equations}

Interiors of stars in equilibrium can be well described by static perfect fluid~\cite{shapiro},
\beq
T^{\mu}_{\nu} = \textrm{diag}(\rho(r), p(r), p(r), p(r)).
\eeq
In this case, the metric can be parameterized
\beq
ds^2 = -e^{A(r)} dt^2 + e^{B(r)} dr^2 + r^2 d\Omega^2.
\eeq
Supplemented by an equation of state, $p = p(\rho)$, there are three functions to be solved, namely the metric functions $A(r)$ and $B(r)$, and the spatial structure of the star $\rho = \rho(r)$.

The continuity equation~(\ref{eq:cont}) reads
\beq
p' = -\frac{A'}{2}(\rho + p). \;
\label{eq:cont_r}
\eeq
The $rr$ and $tt$ components of the field equation~(\ref{eq:field1}) take the forms
\begin{equation}
A' = - \frac{1}{1 + \gamma} \l( \frac{1 - e^B}{r} - \frac{e^B}{F} 8 \pi rp + \frac{\alpha}{r} \r), \:
\label{eq:Aprime}
\end{equation}
\begin{equation}
B' = \frac{1}{1 + \gamma} \l( \frac{1 - e^B}{r} + \frac{e^B}{F} 8 \pi r \rho + \frac{\alpha + \beta}{r} \r), \:
\label{eq:Bprime}
\end{equation}
respectively, as is shown in~\cite{kainulainen}. The variables
\begin{eqnarray}
    \alpha & \equiv & r^2 \left(
            \frac{3}{4}\left(\frac{F'}{F}\right)^2  + \frac{2F'}{rF}
            + \frac{e^B}{2} \left( R - \frac{f}{F} \right)
        \right), \:
\label{eq:alpha} \\
    \beta & \equiv & r^2 \left(
            \frac{F''}{F} - \frac{3}{2}\left(\frac{F'}{F}\right)^2
        \right), \:
\label{eq:beta} \\
     \gamma & \equiv \frac{rF'}{2F}, \:
\label{eq:gamma}
\end{eqnarray}
depend on the first and second derivatives of $F$. The trace equation~(\ref{eq:trace}) yields
\beq
 \frac{F''}{F} = N_1 p'' + N_2 p'^2,
\label{eq:ddF}
\eeq
\beq
 \frac{F'}{F} = N_1 p',
\label{eq:dF}
\eeq
where
\beqa
 N_1 &=& \frac{1}{F} \frac{\partial F}{\partial R} \frac{\partial R}{\partial T} \l(3 - \frac{\partial \rho}{\partial p} \r),
 \: \label{eq:N1} \\
 N_2 &=& \frac{1}{F}  \left[ \frac{\partial^2 F}{\partial R^2} \l(\frac{\partial R}{\partial T} \r)^2 + \frac{\partial F}{\partial R} \frac{\partial^2 R}{\partial T^2} \right] \l( 3 - \frac{\partial \rho}{\partial p} \r)^2 \nonumber \\
 & & {} - \frac{1}{F} \frac{\partial F}{\partial R} \frac{\partial R}{\partial T} \frac{\partial^2 \rho}{\partial p^2} .
 \: \label{eq:N2}
\eeqa
Using the continuity equation~(\ref{eq:cont_r}), its derivative, and the $\theta \theta$ component of the field equation~(\ref{eq:field1}),
\beqa
A'' &=& -A'^2 + \half A' B' - \frac{1}{r}(A' - B') \nonumber \\
& & + \frac{1}{F} \bigg[ \frac{16 \pi}{r^2}e^B T_{\theta \theta} + e^B (f - RF) - 2 F'' \nonumber \\
& & + \frac{3}{2} \frac{F'^2}{F} + (B' - A' - \frac{2}{r}) F' \bigg],
\eeqa
we end up with the equations
\beq
  a p'^2 + b p' + c = 0, \:
\label{eq:pprime}
\eeq
\beqa
     & & p'' = \bigg[ N_2 p'^2 - \half A' \bigg(\frac{\rho' + p'}{\rho + p} - \frac{d}{r} \bigg) \nonumber \\
     & & - \frac{d}{2(1 + \gamma)} \bigg( N_2 p'^2 + \frac{1-e^B}{r^2} + \frac{e^B}{2F}(FR - f) \nonumber \\
     & & + \frac{8 \pi e^B}{F} \rho + \frac{\gamma (4 - 3\gamma)}{r^2} \bigg) + \frac{e^B}{2F} \l( FR - f \r) - \frac{8 \pi e^B}{F} p \nonumber \\
     & & + \frac{\gamma (2 - 3 \gamma)}{r^2} \bigg] \cdot \frac{\rho + p}{1 + N_1 (\rho + p) \big[ \frac{d}{2(1 + \gamma)} - 1 \big] },
     \:
\label{eq:pprimeprime}
\eeqa
where
\beqa
 a &=& rN_1 \Big[ 1 - \frac{3}{4}(\rho + p)N_1 \Big], \:
\label{eq:a} \\
 b &=& 2 \big[ 1 - (\rho + p)N_1 \big], \:
\label{eq:b} \\
 c &=& -(\rho + p) \Big[ \frac{1 - e^B}{r} - \frac{8 \pi re^B}{F} p + \frac{re^B}{2F}(FR - f) \Big] \:
\label{eq:c}
\eeqa
and
\beq
 d = \half r A' + 2 \gamma + 1.
\eeq
The equation~(\ref{eq:pprime}) can be solved by
\beqa
p' = \bigg\{
 \begin{array}{rl}
 \frac{-b \pm \sqrt{b^2 - 4ac}}{2a}, & a \neq 0 \\
 -\frac{c}{b}, & a = 0.  
  \label{eq:depa}
 \end{array}
\eeqa
In general relativity, $a=0$. The "+" branch of Eq.~(\ref{eq:depa}) gives general relativity at the limit $a \rightarrow 0$
\begin{align*}
\lim_{a \rightarrow 0} \frac{-b + \sqrt{b^2 - 4ac}}{2a} = -\frac{c}{b},
\end{align*}
while the "-" branch diverges.

When supplemented by an equation of state $p = p(\rho)$, the equations~(\ref{eq:pprime}) and~(\ref{eq:Bprime}) accompanied by the equation~(\ref{eq:pprimeprime}) form a closed set of differential equations for $p$ and $B$.
\begin{figure}[!t]
	\begin{center}
	\includegraphics[width=8cm]{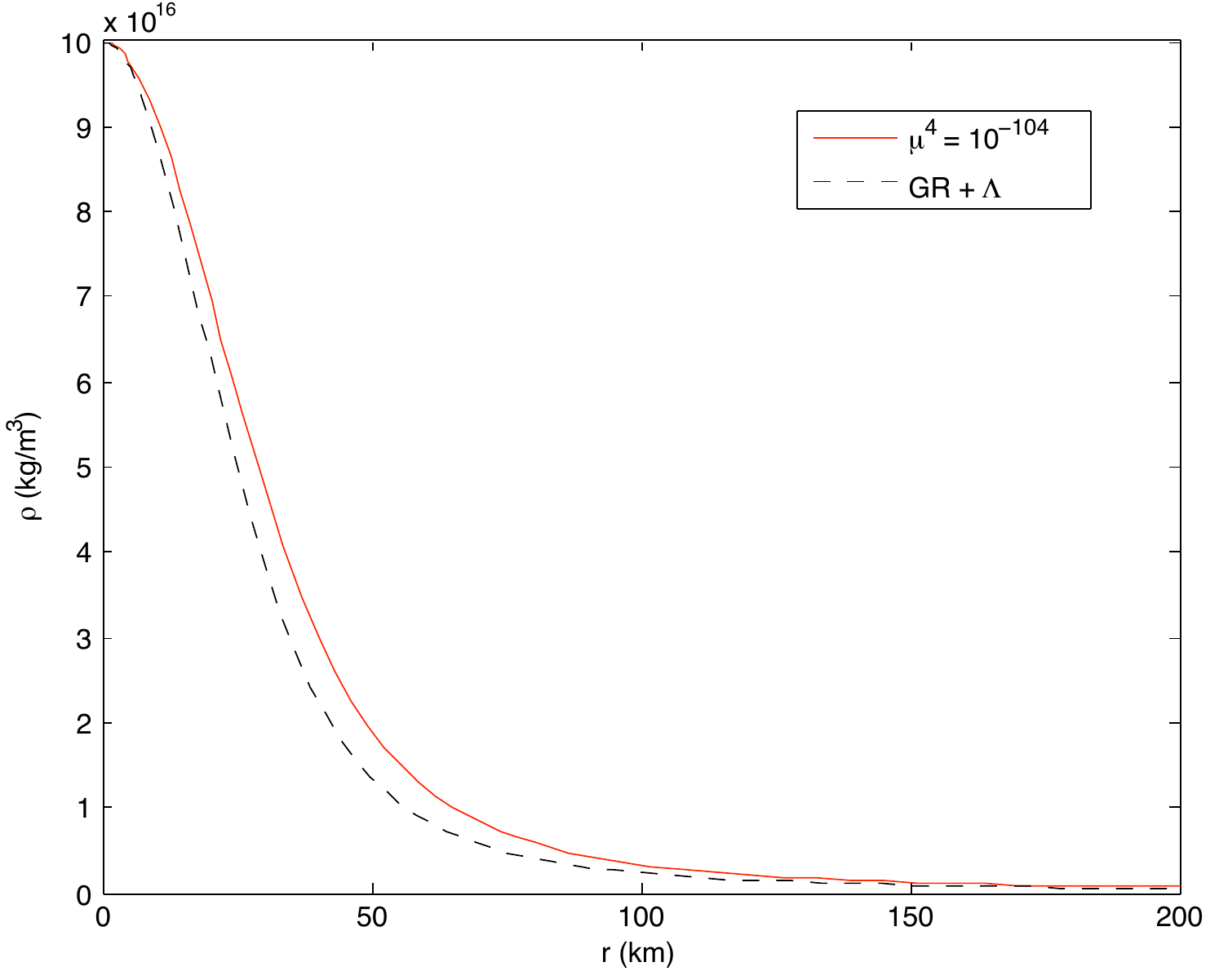}
	\end{center}
	\caption{The energy density profile of a static, spherical object with the central density $10^{17}$ kg/m$^3$ and the equation of state of radiation, $p = \rho /3$, in two different gravity models: $f(R) = R - \mu^4/R$ with $\mu^4 = 10^{-104}$ m$^{-4}$ (red solid line) and general relativity with a cosmological constant $\Lambda = 10^{-52}$ m$^{-2}$ (black dashed line).}
	\label{fig:rad}
\end{figure}

\subsection{The exterior solution in Palatini f(R)}
\label{sec:exterior}

As mentioned above, the trace equation~(\ref{eq:trace}) relates the Ricci scalar to the trace of the energy-momentum tensor algebraically in the Palatini formalism. For instance in the case of $f(R) = R - \mu^4/R$ the trace equation yields
\beq
 R = -4 \pi T \pm \sqrt{(4 \pi T)^2 + 3 \mu^4},
\label{eq:tracemu}
\eeq
while for $f(R) = R + \alpha R^2$ the trace equation is the same as in general relativity, $R = -8 \pi T$. By the exterior metric of the star we imply the vacuum solution $T_{\mu \nu} = 0$, the modified Schwarzschild metric. Since, in that case, $R$ and thus $f$ and $F$ are constants, the exterior metric is easily and uniquely solved from~(\ref{eq:Aprime}) and~(\ref{eq:Bprime}):
\begin{equation}
A = \textrm{ln} \l( 1 - \frac{2M}{r} - \frac{r^2}{3} \Lambda_{eff} \r) \:
\label{eq:AprimeExt}
\end{equation}
\begin{equation}
B = -\textrm{ln} \l( 1 - \frac{2M}{r} - \frac{r^2}{3} \Lambda_{eff} \r), \:
\label{eq:BprimeExt}
\end{equation}
where $M$, the Schwarzschild mass, and
\beq
\Lambda_{eff} = \half \l( R - \frac{f}{F} \r) + \frac{\Lambda}{F}
\label{eq:effLambda}
\eeq
are constants. For the model $f(R) = R - \mu^4/R$,
\beq
\Lambda_{eff} = \frac{3}{4} \l( \frac{\mu^2}{\sqrt{3}} + \Lambda \r).
\label{eq:effmu4}
\eeq
Although in general one cannot conclude the nature of the interior solution ($\rho \neq 0$) from the trace equation alone, it can be done in the case of $p = \rho /3$, which is the equation of state of radiation and ultrarelativistic matter, and which also has $T = 0$. The case is particularly interesting since as the matter in stars gets more and more relativistic, the equation of state gets closer to the ultrarelativistic one. The equations~(\ref{eq:Aprime}) and~(\ref{eq:Bprime})
\beq
A' = -\left[ \frac{1 - e^B}{r} - e^B r \l( \frac{8 \pi}{F}p + \Lambda_{eff} \r) \right]
\eeq
\beq
B' = \frac{1 - e^B}{r} + e^B r \l( \frac{8 \pi}{F}\rho + \Lambda_{eff} \r)
\eeq
are the same as in general relativity except for the constant coefficient $F$ and the effective cosmological constant~(\ref{eq:effLambda}). In the case of $f(R) = R - \mu^4/R$, $F = 4/3$ and $\Lambda_{eff}$ is given by Eq.~(\ref{eq:effmu4}). Hence, no matter how small the parameter $\mu^4$ is, there is a visible difference to GR (Fig.~(\ref{fig:rad})). In the case $f(R) = R + \alpha R^2$, $F = 1$ and $\Lambda_{eff} = \Lambda$ and there is no difference to GR. As shown in the following, this is actually opposite to what happens before the limit of the equation of state of radiation is reached: the $\mu^4$ case implies practically no changes to the structure of the stars while the $\alpha$ model yields a notable change already at relatively small parameter values.

%
%

\section{Fermi stars in f(R) models}

\subsection{Fermi gas equation of state}

By the Pauli exclusion principle, the pressure of a gas of fermions is non-zero even at zero temperature. After burning all the nuclear fuel that sustains thermal pressure, a star begins to collapse due to gravity. Ultimately the degeneracy pressure of electrons or neutrons (or, hypothetically, quarks) takes over, stabilizing the star to a white dwarf or a neutron star (or a quark star), respectively. If the pressure of the matter states the star goes through while collapsing denser and denser never prevails over gravity, the star is believed to form a black hole.

In the completely degenerate case, the fermions occupy the lowest momenta states up to the Fermi momentum
\beq
p_F = \l( \frac{3 h^3 \rho_0}{8 \pi m} \r)^{1/3},
\eeq
where $\rho_0$ is the rest frame energy density of the fermions of mass $m$. The pressure and the energy density of the spin-$\half$ Fermi gas are easily integrated analytically~\cite{oppenheimer},
\beqa
p &=& \frac{8 \pi}{3 h^3} \int_0^{p_F} dp p^3 \frac{dE}{dp} \nonumber \\
&=& \frac{1}{3} K \left[ \textrm{sinh}(t) - 8 \textrm{sinh} \l( \half t \r) + 3t \right],
\label{eq:Fermi_p}
\eeqa
\beq
\rho = \frac{8 \pi}{h^3} \int_0^{p_F} dp p^2 E = K \left[ \textrm{sinh}(t) - t \right],
\label{eq:Fermi_rho}
\eeq
where $E = m \l( \sqrt{1 + \frac{p^2}{m^2}} - 1 \r)$, $K = \frac{\pi m^4}{4 h^3}$ and
\beq
t = 4 \textrm{ln} \l( \frac{p_F}{m} + \sqrt{1 + \frac{p_F^2}{m^2}} \r).
\eeq
\begin{figure}[!t]
	\begin{center}
	\includegraphics[width=8cm]{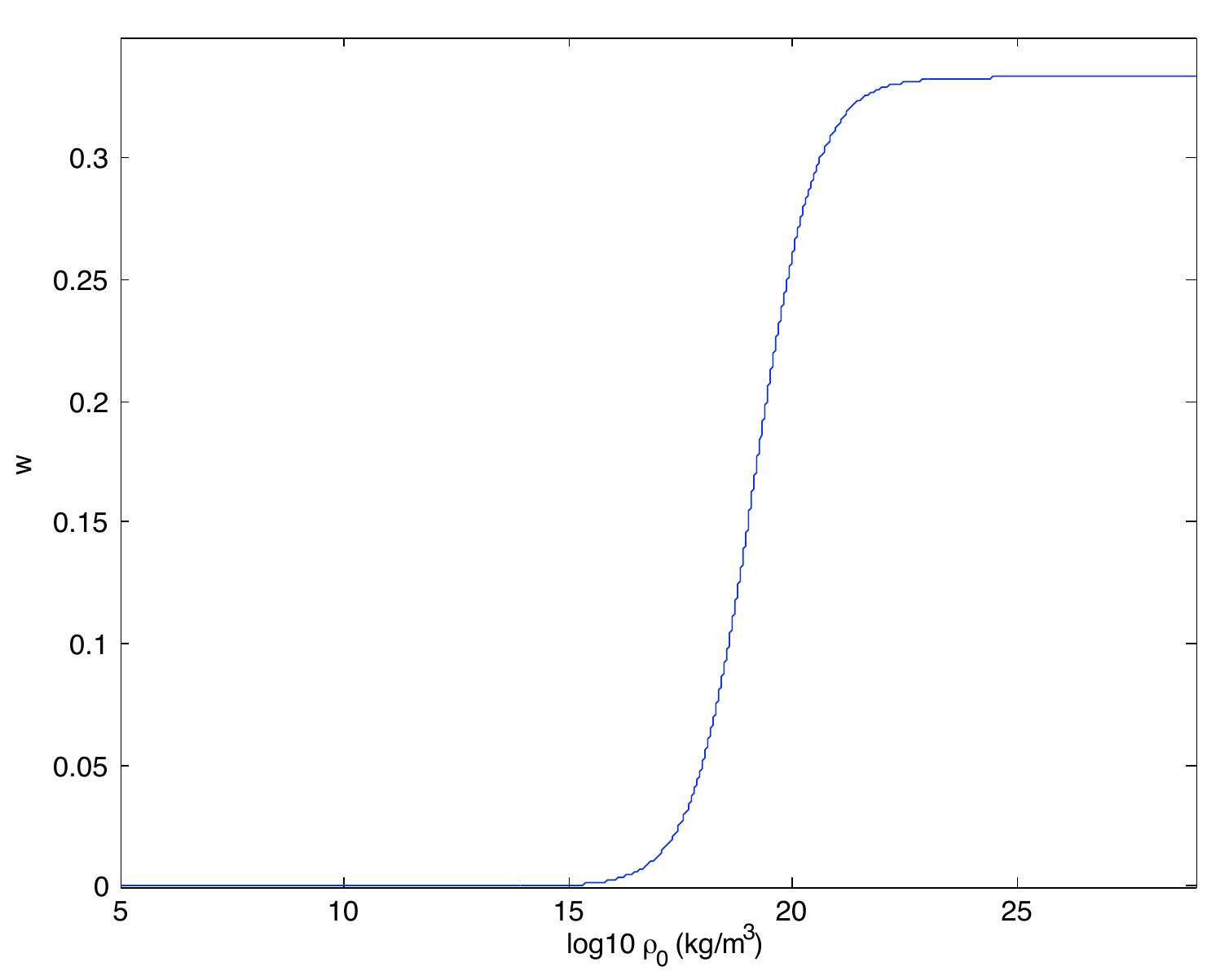}
	\end{center}
	\caption{The equation of state parameter of Fermi gas $w$ (Eq. ~(\ref{eq:w_Fermi})) as a function of the rest energy density $\rho_0$.}
	\label{fig:eos}
\end{figure}

Eq.~(\ref{eq:Fermi_p}) has the asymptotic forms
\beq
p = \frac{1}{3} K \bigg\{
 \begin{array}{rl}
  \frac{8}{5} x^5 - \frac{4}{7} x^7 + ... ,& x \to 0  \\
  2 x^4 - 3 x^2 + ... ,& x \to \infty,
 \end{array}
\eeq
where $x = p_F/m$. When only the first terms in these expansions are retained, we are left with the familiar polytropes of non-relativistic $(x \to 0)$ and ultra-relativistic $(x \to \infty)$ Fermi gas
\beq
p = K_1 \rho_0^{5/3},
\label{eq:NR_eos}
\eeq
\beq
p = K_2 \rho_0^{4/3},
\eeq
respectively, with 
\beq
K_1 = \l( \frac{3}{\pi} \r)^{2/3} \frac{h^2}{20 m (\mu M)^{5/3}}
\label{eq:K1} 
\eeq
and 
\beq
K_2 = \l( \frac{3}{\pi} \r)^{1/3} \frac{hc}{8 (\mu M)^{4/3}}. 
\eeq
One should read $\mu = 1$ and $M = m$ for neutron stars, while typically for white dwarfs $\mu = 2$ (number of nucleons per one electron) and $M = m_p$ (the average nucleon mass)~\cite{chandra}. \\

\begin{figure}[!t]
	\begin{center}
	\includegraphics[width=8cm]{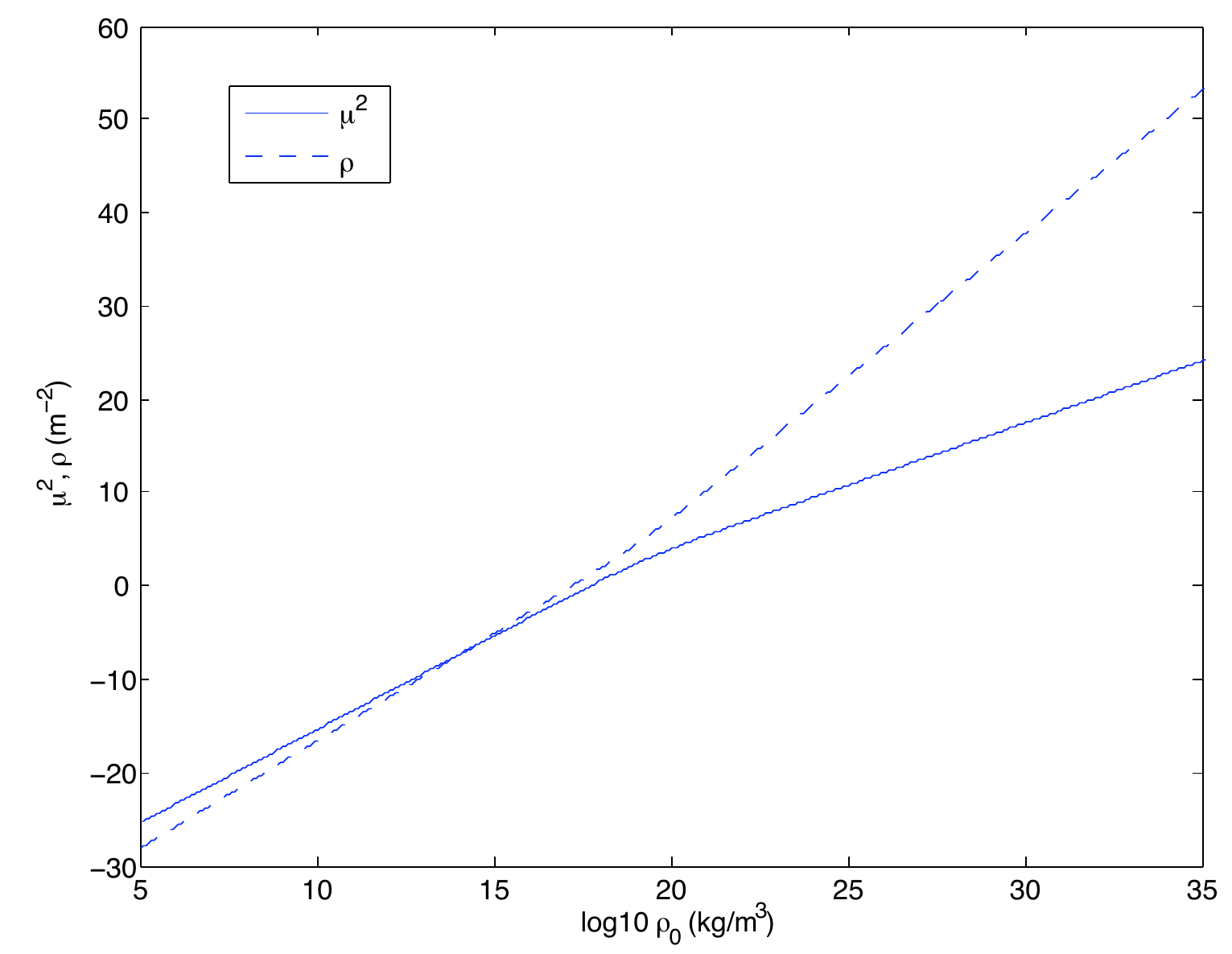}
	\end{center}
	\caption{The lower limit of the inequality~(\ref{eq:ineqmu}), $\mu^2 = (1 - 3w) \rho$ (and the total energy density $\rho$) as a function of the rest energy density $\rho_0$ in the case of a pure degenerate neutron gas.}
	\label{fig:mu2}
\end{figure}

The equation of state parameter of the pure Fermi gas (that is, $\rho$ including only the degenerate fermions $-$ in white dwarfs the plasma is composed of electrons \emph{and} nuclei),
\beq
w \equiv \frac{p}{\rho} = \frac{1}{3}\frac{\textrm{sinh}(t) - 8 \textrm{sinh} \l( \half t \r) + 3t}{\textrm{sinh}(t) - t},
\label{eq:w_Fermi}
\eeq
approaches $1/3$ at high (rest) energy densities as shown in Fig.~(\ref{fig:eos}). As reasoned in Section~(\ref{sec:exterior}), we would expect to see a change in the star profiles even at very low parameter $\mu^4$ values if the equation of state is close to $1/3$. The question is, \emph{how} close? It turns out that the Fermi equation of state parameter~(\ref{eq:w_Fermi}) is close enough to $1/3$ for uninterestingly high values of $\mu^4$. One way to see this is to compare the $\mu^2$ and $T (= (3w - 1)\rho)$ terms in Eq.~(\ref{eq:tracemu}): at low energies $w << 1/3$ and $-T >> \mu^2$ (for any natural value of $\mu^2$), but as we reach higher energies and $w \to 1/3$, the two terms start to approach each other. If
\beq
\mu^2 \gtrsim (1 - 3w) \rho,
\label{eq:ineqmu}
\eeq
definitely there should appear a change in the star profile. In Figure~(\ref{fig:mu2}) we have plotted this lower limit on $\mu^2$ in the case of the Fermi equation of state. Comparing the plot $\mu^2(\rho_0)$ to the plot of $\rho(\rho_0)$ we see that the feature $w \to 1/3$ does push the limiting $\mu^2$ down by several magnitudes, but nevertheless not even close to any physically interesting values. The value $\mu^4_{DE} = 10^{-104} $m$^{-4}$ that yields dark energy is simply so tiny that we cannot see any difference with respect to general relativity in the case of compact stars.

\subsection{The polytropic divergence}

The viability of Palatini $f(R)$ gravity has been questioned by Barausse \emph{et al}~\cite{Barausse:2007pn} because the factor $F''$ in the field equations diverge at $p \to 0$ (the surface of the star) \emph{for polytropes} $p = K \rho_0^{\Gamma}$ with $\Gamma > 3/2$. The Fermi equation of state~(\ref{eq:Fermi_p}) approaches the form~(\ref{eq:NR_eos}) $-$ that is $\Gamma = 5/3$ $-$ at the edge of the star, so the divergence might seem worrying. While it truly inflicts a qualitative difference between general relativity and Palatini $f(R)$ theories, one should inspect the situation quantitatively. The diverging term is given by Eq.~(\ref{eq:ddF}), and due to Eqs.~(\ref{eq:pprimeprime}) and~(\ref{eq:pprime}), it is proportional to
\beq
N_2 p'^2 \propto  C_{f(R)}(T) \l( p + \rho \r)^2 \frac{\partial^2 \rho}{\partial p^2} \propto C_{f(R)} \rho_0^{3 - 2\Gamma}
\eeq
at the limit $p \to 0$, where $C_{f(R)} \equiv \frac{1}{F} \frac{\partial F}{\partial R} \frac{\partial R}{\partial T}$. As is noticed in~\cite{Olmo:2008pv}, in the case of $f(R) = R + \alpha R^2$ the factor
\beq
C_{f_{\alpha}} = \frac{1}{T - \frac{1}{16 \pi \alpha}} \propto \alpha, \qquad (T << 1/\alpha)
\eeq
suppresses the divergence to very low densities. The density scale at which the divergence becomes of order unity is 
\beq
\rho_{S, \alpha} \sim (K^2/\alpha)^{1/(3 - 2\Gamma)}.
\label{eq:rhoSalpha}
\eeq
The maximum value $-$ $\rho_{S, \alpha} \sim 10^{-46}$ kg/m$^3$ $-$ that will be encountered in this paper is the case of $\alpha = 10^{-16}$ m$^2$ and $K = K_1$ (Eq.~(\ref{eq:K1})) for neutrons.

In the case of $f(R) = R - \mu^4/R$,
\beqa
C_{f_{\mu}} &=& \frac{8 \pi}{\mu^2} \frac{1}{\sqrt{\tilde{T}^2 + 3} \left[(\sqrt{\tilde{T}^2 + 3} - \tilde{T})^2 + 1 \right]} \nonumber \\
&\to& \frac{1}{32 \pi^2} \frac{\mu^4}{|T|^3}, \qquad (|T| >> \mu^2)
\eeqa
where $\tilde{T} \equiv 4 \pi T / \mu^2$, and the density scale at which the divergence becomes of order unity is
\beq
\rho_{S, \mu} \sim (K^2/\mu^4)^{-1/2\Gamma}.
\label{eq:rhoSmu}
\eeq
In the case of $\mu^4 = 10^{-104}$ m$^{-4}$ and $K = K_1$ (Eq.~(\ref{eq:K1})) for neutrons, $\rho_{S, \mu} \sim 10^{-7}$ kg/m$^3$.

Formally the limit $p_F \to 0$ can be taken, and $p$ as well as $\rho$ and $\rho_0$ approach smoothly zero. Physically, however, the Fermi gas equation of state breaks down already at a higher density inside stars. White dwarfs have a gas envelope and neutron stars an outer crust, where the matter is not degenerate. The polytropic divergence does not practically restrict the $f(R)$ model parameters much: if we insist the validity of Eq.~(\ref{eq:NR_eos}) until $\rho_s < 10^3 \frac{kg}{m^3}$ (the density of water), the parameters of models $f(R) = R - \mu^4/R$ and $f(R) = R + \alpha R^2$ must take values $\mu^4 \lesssim 10^{-71}$ and $\alpha \lesssim 1$. These upper bounds are so large that they are already excluded by observation~\cite{Kagramanova:2006ax}. 

In this paper we have used the Fermi equation of state, Eqs.~(\ref{eq:Fermi_p}) and~(\ref{eq:Fermi_rho}), to track profiles of stars down to as low densities as the efficiency (accuracy vs. time) of numerics has admitted. The divergences arise as approximated by Eqs.~(\ref{eq:rhoSalpha}) and~(\ref{eq:rhoSmu}). These densities are so small in the models we have chosen to present in this paper that we entitle ourselves to cut the star profiles well before the divergences arise.

\subsection{Results: density profiles and stability curves}

In general relativity, we are used to the so-called mass parameter $m(r) \equiv r \l(1 - e^{-B(r)} \r) /2$, which is nothing but a reparameterization of the metric function $B(r)$. In terms of the mass parameter, the field equation~(\ref{eq:Bprime}) takes the form
\beqa
m' = \frac{1}{1 + rF'/2F} \left[ \frac{1}{F} 4 \pi r^2 \rho - \frac{m}{r} + \frac{r^2}{4} \l(R - \frac{f}{F} \r) \right. \nonumber \\
 \left. + \frac{r^2}{r} \l( 1 - \frac{2m}{r} \r) \l(\frac{F''}{F} - \frac{3}{4}\l(\frac{F'}{F}\r)^2 + \frac{2F'}{rF} \r) \right] + \frac{m}{r},
\label{eq:mprime}
\eeqa
which in the case of general relativity separates to the simple form $m(r) = \int_0^r dr 4\pi r^2 \rho(r) + r^3 \Lambda/6$. The interior solution is smoothly connected to the exterior Schwarzschild solution by identifying $M \equiv m(r_f) - r_f^3 \Lambda_{eff}/6$, where $r_f$ is the radius of the object. When we speak of the mass of a star, we mean the Schwarzschild mass.

The total energy of a static, spherical object has to be defined by the Schwarzschild mass $M$, since it completely governs the geometry outside the matter (except for the (effective) cosmological constant). The equilibria for a number of particles are the configurations that extremize the Schwarzschild mass, and these configurations are indeed what we get by solving the (modified) Einstein field equations~\cite{htww}.

Although abstruse, the equation~(\ref{eq:pprime}) for $dp[\rho(r)]/dr$ (accepting only the "proper" solution, which yields general relativity as a limit) and the equation~(\ref{eq:mprime}) for $dm(r)/dr$ are first order differential equations with respect to $r$, and the solution is uniquely defined by the initial values $m(0) = 0$ (equals to setting $B(0) = 0$) and $\rho(0) = \rho_C$ like in GR. Thus the central density $\rho_C$ may be safely used as the parameter labeling the different equilibrium configurations $-$ and, moreover, to serve together with $m(0)$ (or $B(0)$) as the initial values in solving the field equations. At the same time, for instance, the total particle number or the radius of the object do not uniquely define the solution. 

\begin{figure}[t]
	\begin{center}
	\includegraphics[width=8cm]{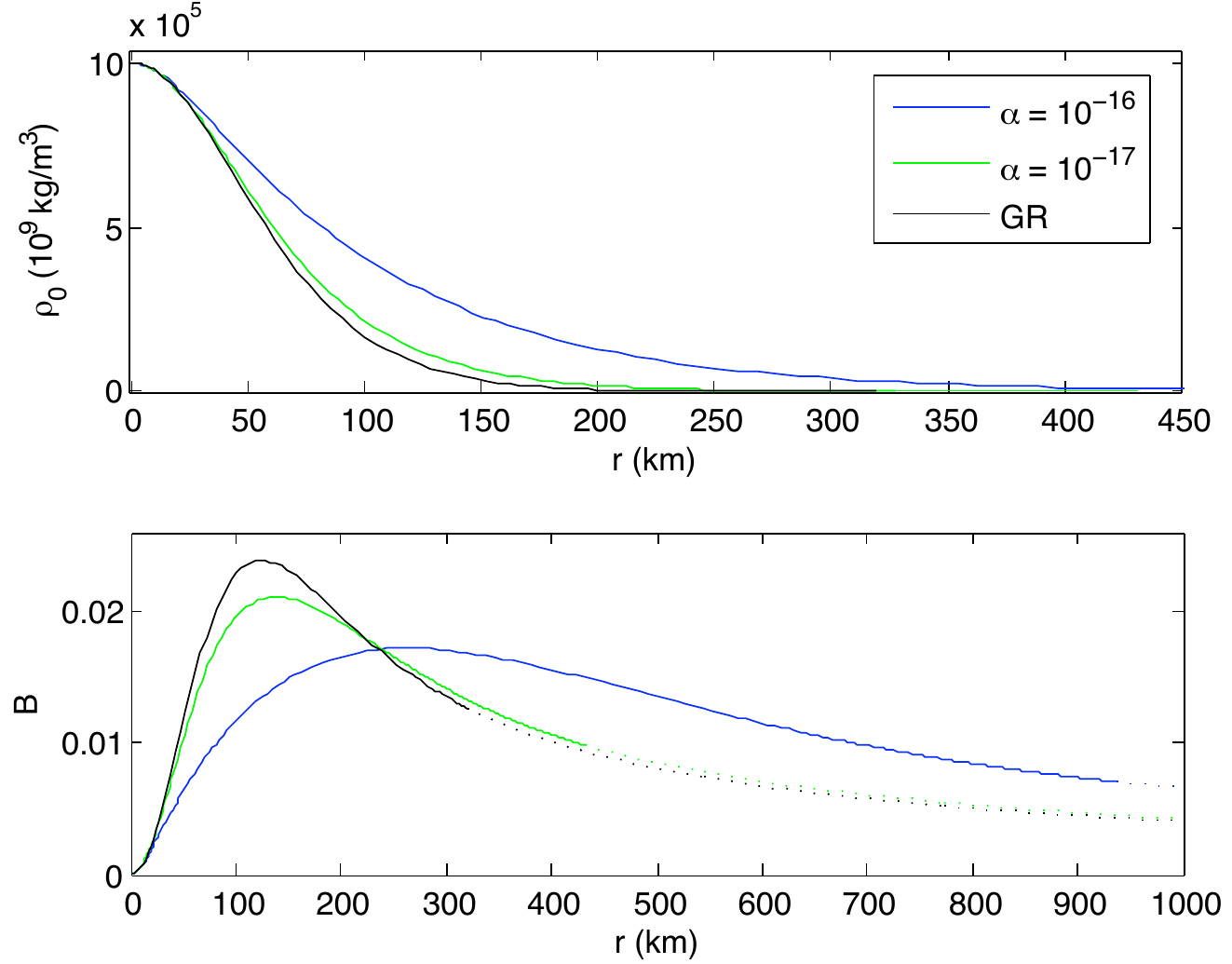}
	\end{center}
	\caption{The rest energy density profiles $\rho_0(r)$ of white dwarfs with the same central density ($10^{15}$ kg/m$^3$) in three different models of gravity: $f(R) = R + \alpha R^2$ (blue curve $\alpha = 10^{-16}$ m$^2$, green curve $\alpha = 10^{-17}$ m$^2$) and general relativity (black curve). The metric function $B(r)$ is presented for the same models (solid line is for the interior solution, dashed line for the continued exterior (Schwarzschild) solution). The Schwarzschild masses of the stars are $2.21$, $1.43$ and $1.36 M_{Sun}$, respectively.}
	\label{fig:wd_profs}
\end{figure}

\begin{figure}[t]
	\begin{center}
	\includegraphics[width=8cm]{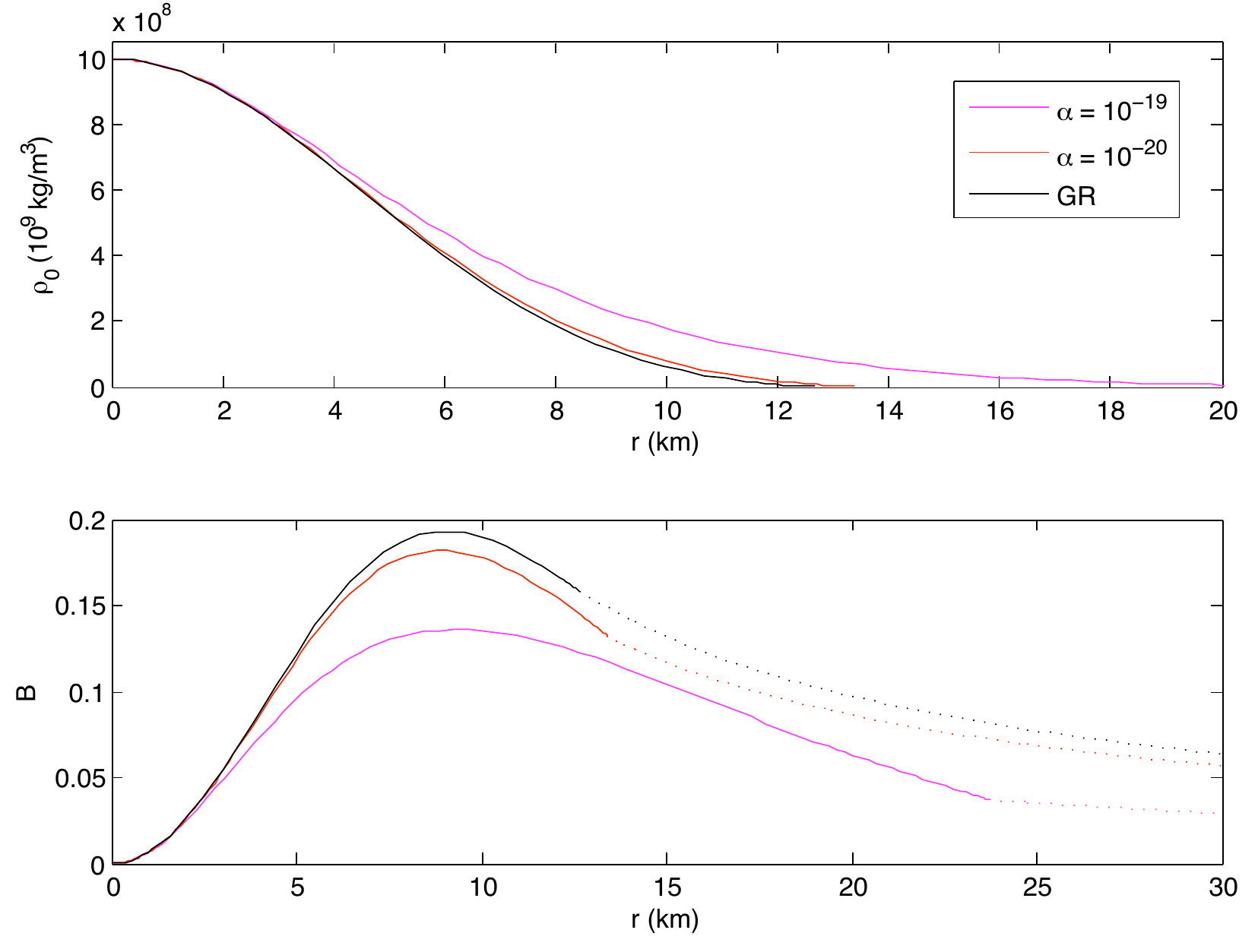}
	\end{center}
	\caption{The rest energy density profiles $\rho_0(r)$ of neutron stars with the same central density ($10^{18}$ kg/m$^3$) in three different models of gravity: $f(R) = R + \alpha R^2$ (magenta curve $\alpha = 10^{-19}$ m$^2$, red curve $\alpha = 10^{-20}$ m$^2$) and general relativity (black curve). The metric function $B(r)$ is presented for the same models (solid line is for the interior solution, dashed line for the continued exterior (Schwarzschild) solution). The Schwarzschild masses of the stars are $0.239$, $0.554$ and $0.625 M_{Sun}$, respectively.}
	\label{fig:neutron_prof1}
\end{figure}

\begin{figure}[t]
	\begin{center}
	\includegraphics[width=8cm]{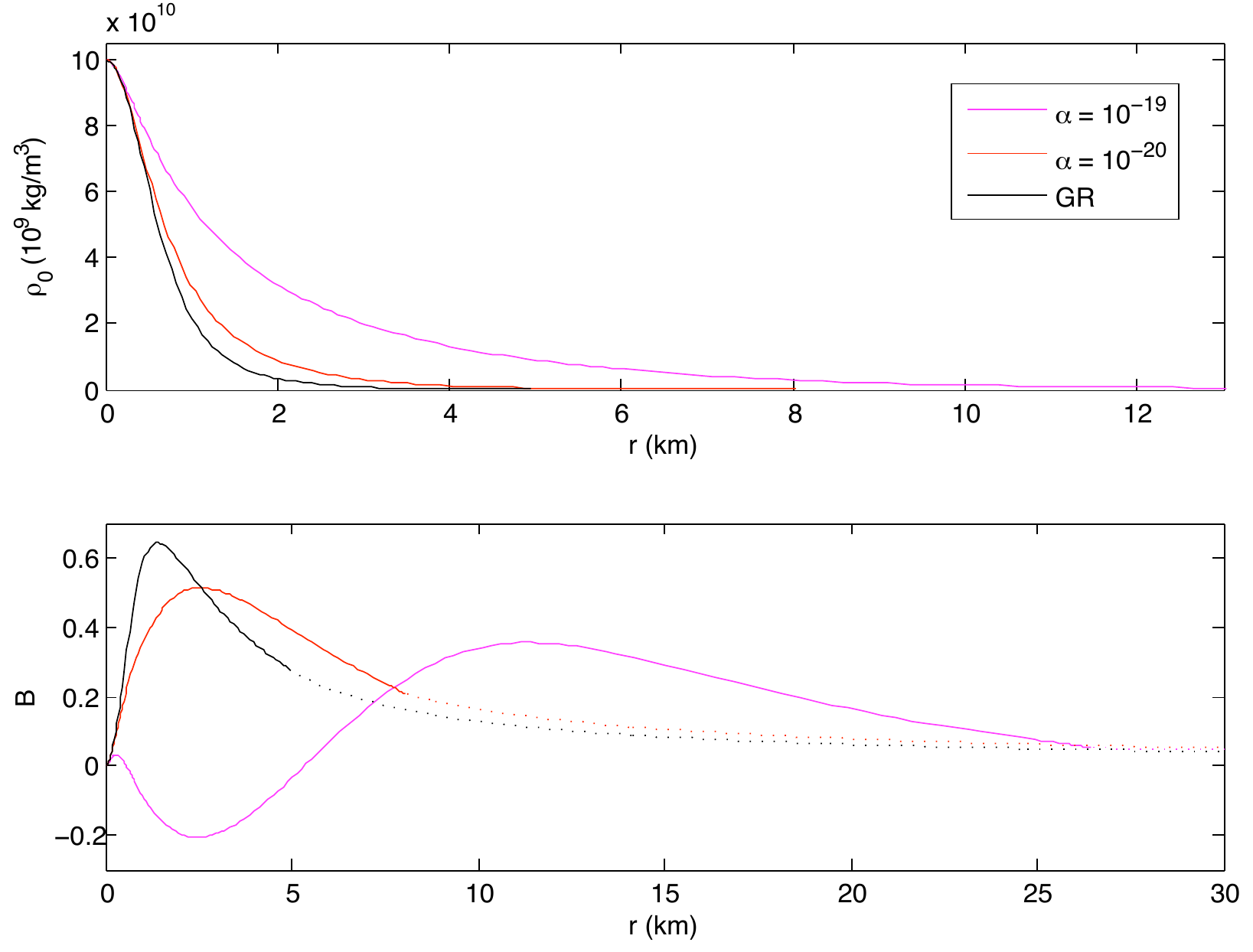}
	\end{center}
	\caption{The rest energy density profiles $\rho_0(r)$ of neutron stars with the same central density ($10^{20}$ kg/m$^3$) in three different models of gravity: $f(R) = R + \alpha R^2$ (magenta curve $\alpha = 10^{-19}$ m$^2$, red curve $\alpha = 10^{-20}$ m$^2$) and general relativity (black curve). The metric function $B(r)$ is presented for the same models (solid line is for the interior solution, dashed line for the continued exterior (Schwarzschild) solution). The Schwarzschild masses of the stars are $0.419$, $0.494$ and $0.406 M_{Sun}$, respectively.}
	\label{fig:neutron_prof2}
\end{figure}

In Figure~(\ref{fig:wd_profs}) we present, as examples of our white dwarfs, the rest energy density profiles and the behavior of the metric function $B(r)$ in three different cases: in Palatini $f(R) = R + \alpha R^2$ with $\alpha = 10^{-16}$ m$^2$ and $10^{-17}$ m$^2$, and in general relativity, all starting with the same central density ($\rho_C = 10^{15}$ kg/m$^3$) but ending with different values of the Schwarzschild mass ($2.21$, $1.43$ and $1.36$ Solar mass, respectively). The larger $\alpha$ is, the less centrally peaked the profiles seem to become: in this region, gravity is weaker compared to general relativity. 

\begin{figure}[t]
	\begin{center}
	\includegraphics[width=8cm]{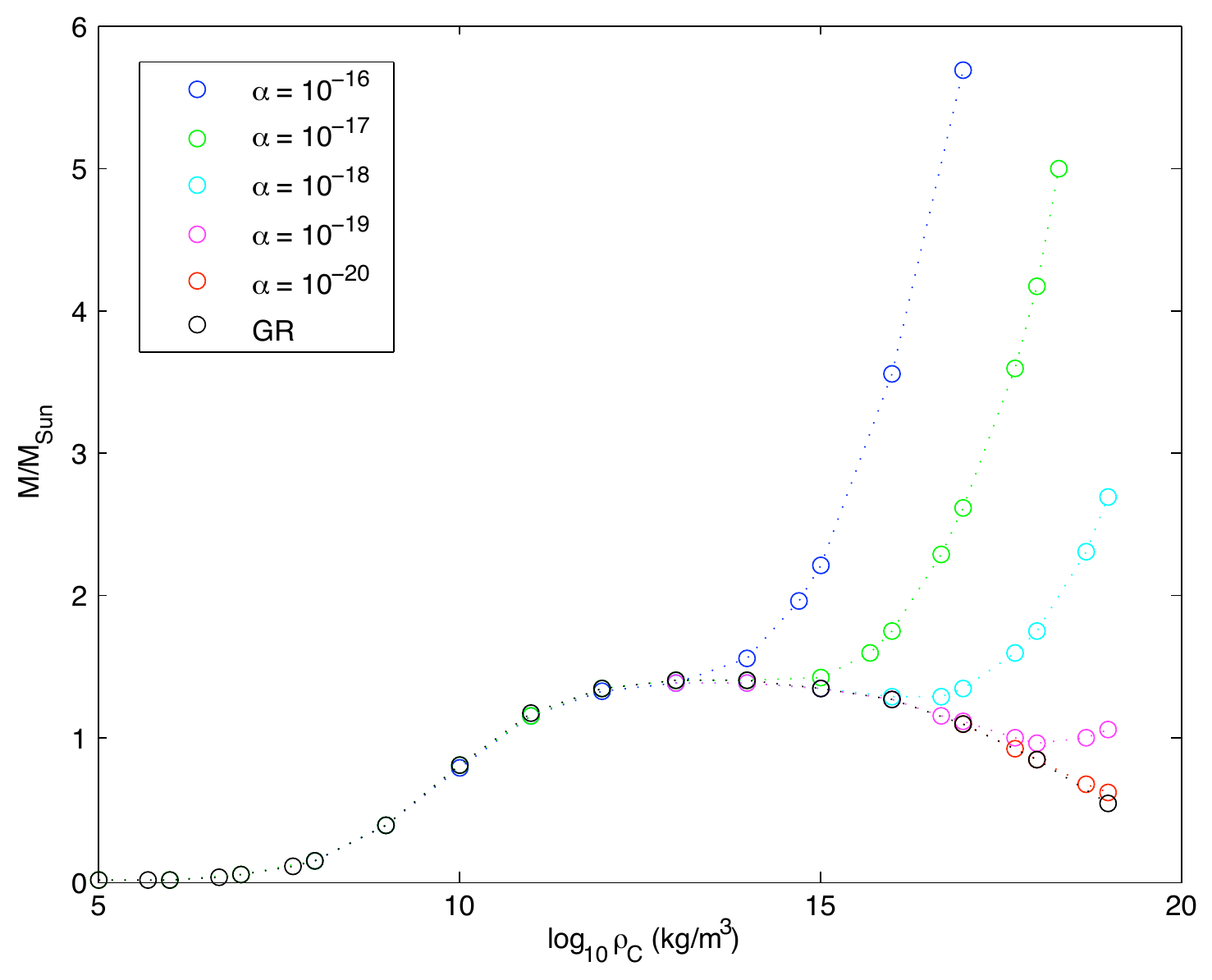}
	\end{center}
	\caption{The Schwarzschild mass to central density curves of white dwarfs in different models of gravity: $f(R) = R + \alpha R^2$ (blue curve $\alpha = 10^{-16}$ m$^2$, green curve $\alpha = 10^{-17}$ m$^2$, cyan curve $\alpha = 10^{-18}$ m$^2$, magenta curve $\alpha = 10^{-19}$ m$^2$, red curve $\alpha = 10^{-20}$ m$^2$ and general relativity (black curve). }
	\label{fig:wd_curve}
\end{figure}

\begin{figure}[t]
	\begin{center}
	\includegraphics[width=8cm]{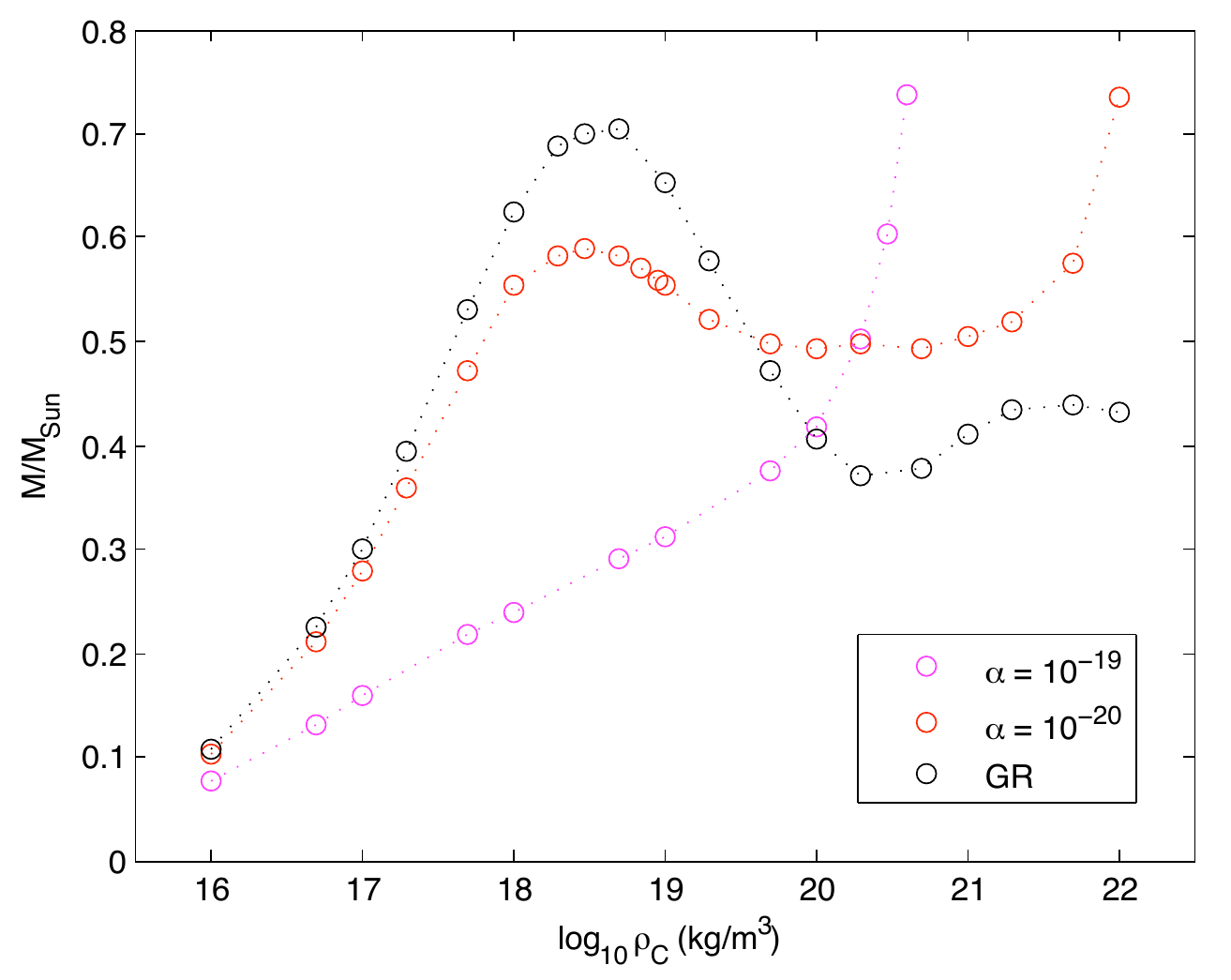}
	\end{center}
	\caption{The Schwarzschild mass to central density curves of neutron stars in three different models of gravity: $f(R) = R + \alpha R^2$ (magenta curve $\alpha = 10^{-19}$ m$^2$, red curve $\alpha = 10^{-20}$ m$^2$ and general relativity (black curve). }
	\label{fig:neutron_curve}
\end{figure}

The dependence of the Schwarzschild mass $M$ on the central density $\rho_C$ for different $f(R)$ models is presented in Figures~(\ref{fig:wd_curve}) and~(\ref{fig:neutron_curve}) for white dwarfs and neutron stars, respectively. In general relativity, the first maxima of the plots are known as the Chandrasekhar limit and the Tolman-Oppenheimer-Volkoff (TOV) limit, the maximum white dwarf mass and the maximum neutron star mass, respectively. In general, the stationary points $dM(\rho_C)/d\rho_C = 0$ are called \emph{turning-points}, and they \emph{may} signal a change from stability to instability, or vice versa.

In the case of static, spherically symmetric solutions, the stability analysis is covered by examining small radial perturbations of the equilibrium configurations. The equations for the frequencies $\omega$ of oscillations of the radial displacements $\xi(r, t) = \xi(r)e^{i \omega t}$, the acoustical modes, are derived in the case of general relativity for example in~\cite{htww}. It can be shown that the turning points signal changes in the stability of different acoustical modes. The frequencies of the acoustical modes form an infinite discrete sequence
\beq
\omega_0^2 < \omega_1^2 < \omega_2^2 < ...,
\label{eq:modes}
\eeq
and depending on the sequence of the turning points stability ($\omega_0^2 > 0$) or instability ($\omega_0^2 < 0$) of the configuration is implied. The logic, which mode changes at which turning point, is straightforward: As we increase the central density, the radius $R$ either decreases or increases depending on the dominating mode in question. In the former case, a mode with an even number of nods dominates and changes stability at the turning point, while in the latter case the mode that changes stability must have odd nods~\cite{shapiro}. Which even/odd mode is in order to change follows from the property~(\ref{eq:modes}).

\begin{figure}[!b]
	\begin{center}
	\includegraphics[width=8cm]{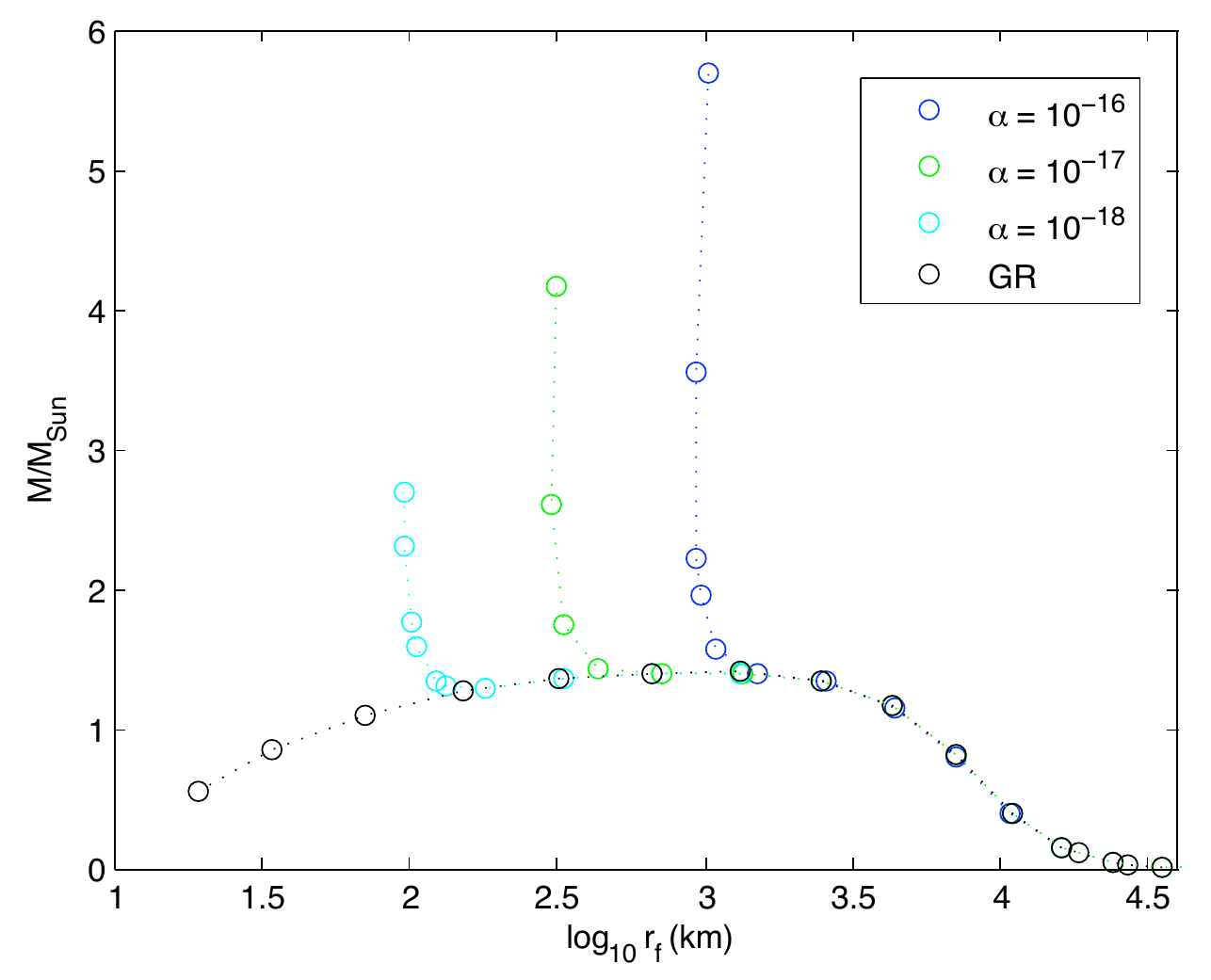}
	\end{center}
	\caption{The Schwarzschild mass to radius of the star curves of white dwarfs in different models of gravity: $f(R) = R + \alpha R^2$ (blue curve $\alpha = 10^{-16}$ m$^2$, green curve $\alpha = 10^{-17}$ m$^2$, cyan curve $\alpha = 10^{-18}$ m$^2$ and general relativity (black curve). }
	\label{fig:wd_M_R}
\end{figure}

\begin{figure}[!b]
	\begin{center}
	\includegraphics[width=8cm]{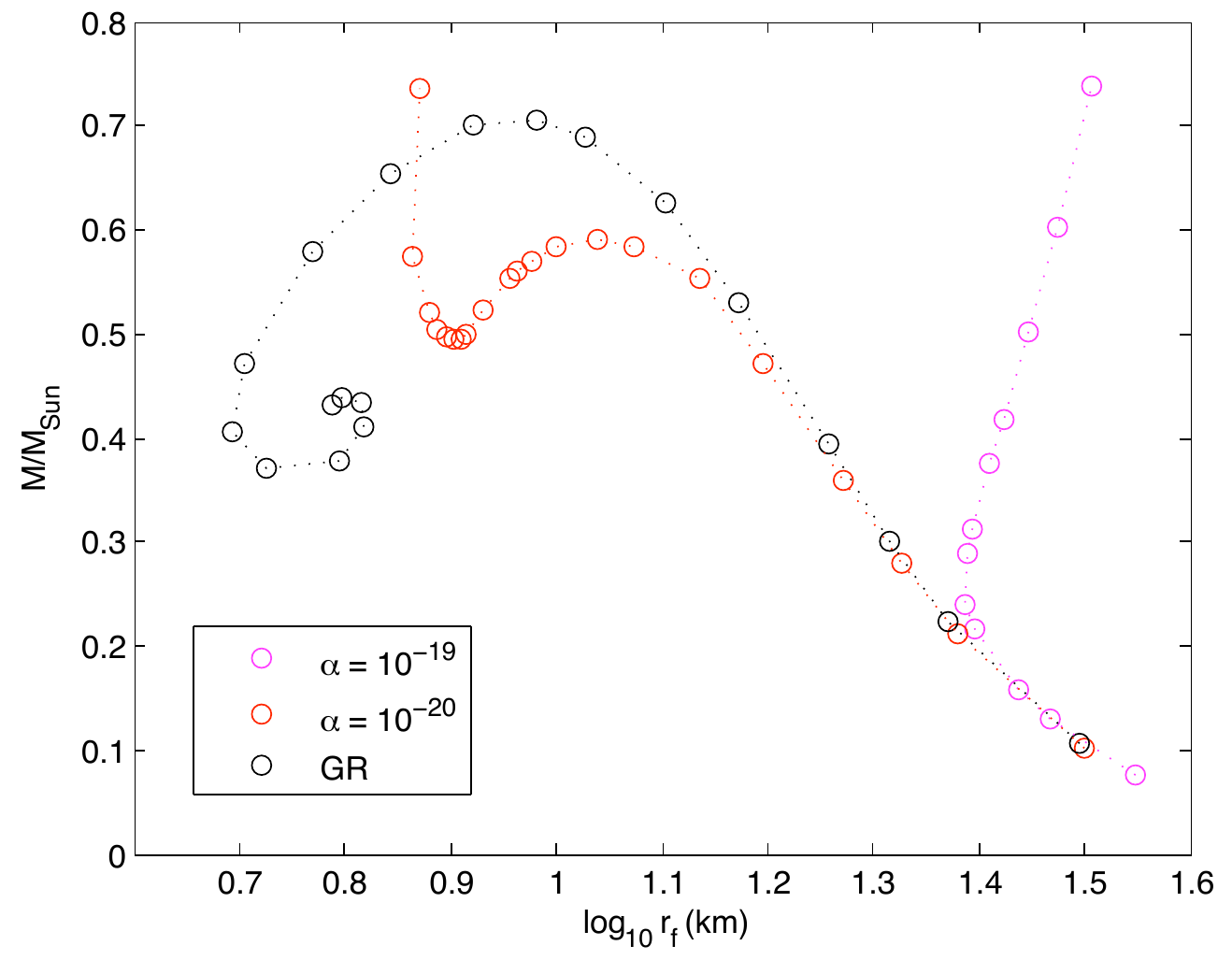}
	\end{center}
	\caption{The Schwarzschild mass to radius of the star curves of neutron stars in three different models of gravity: $f(R) = R + \alpha R^2$ (magenta curve $\alpha = 10^{-19}$ m$^2$, red curve $\alpha = 10^{-20}$ m$^2$ and general relativity (black curve). }
	\label{fig:n_M_R}
\end{figure}

The key result is that the changes in stability can be traced directly from the $M$ versus $r_f$ plot since $dM/dr_f = dM/d\rho_C \cdot d\rho_C/dr_f$: for instance, if all modes are stable at lower central densities, a counterclockwise bend in the $M(r_f)$ plot at a critical point manifests an onset of instability with increasing $\rho_C$ (cf. the Chandrasekhar and TOV limits). The behavior of $dM/dr_f$ in different $f(R)$ models is shown in Figures~(\ref{fig:wd_M_R}) and~(\ref{fig:n_M_R}).

From the white dwarf curves, Fig.~(\ref{fig:wd_curve}) and~(\ref{fig:wd_M_R}), we immediately take notice of the disappearance of Chandrasekhar limit at $\alpha \gtrsim 10^{-16}$ m$^2$. For lower values of $\alpha$ the instability is set on at the Chandrasekhar limit, but it seems to be followed by a region of stable solutions. However, the neutronization of the plasma is expected to begin relatively soon after the "Chandrasekhar critical density" $-$ or even before it, depending on the composition of the star $-$ and to be an even more important factor in the collapse of white dwarfs than the instability due to gravity. 

Our stability curves for white dwarfs were plotted using the equation of state of degenerate electrons alone, that is, not taking the onset of neutronization into account. It is doubtful that the stability regions beyond the neutronization threshold $\rho_C \sim 10^{15}$ kg/m$^3$ would act as barriers against collapsing. Nonetheless, since the Chandrasekhar limit appears to be missing completely for $\alpha \gtrsim 10^{-16}$ m$^2$, the instability is to be triggered by the neutronization alone at a considerably higher $M$ compared to general relativity. For $\alpha \sim 10^{-17}$ m$^2$, the stability region begins slightly before the neutronization threshold, and it is an open question whether condensing taking place between it and the Chandrasekhar limit is enough to set off nuclear reactions (and the explosion of the star). If it were not the case, the white dwarf needs to keep absorbing matter from e.g. a binary partner until the nuclear reactions start taking place at a higher value of the Schwarzschild mass.

The stability curve $M(\rho_C)$ for neutron stars shown in Fig.~(\ref{fig:neutron_curve}) goes through a drastic change between $\alpha = 10^{-19}$ m$^2$ and $10^{-20}$ m$^2$. The Tolman-Oppenheimer-Volkov limit exists for $\alpha \lesssim 10^{-20}$ m$^2$, but it is located at a lower Schwarzschild mass and at a lower central density. Moreover, the instability region subsequent to the TOV turning point seems to be followed by a stability region unlike in general relativity. For $\alpha \gtrsim 10^{-19}$ m$^2$, all solutions are stable and the TOV limit does not exist. 

Qualitatively, the behavior of the stability curve in the case of neutron stars is not as easy to understand as in the case of white dwarfs. From the Figures~(\ref{fig:neutron_prof1}) and~(\ref{fig:neutron_prof2}) we perceive that the density profiles of our neutron stars are the less centrally peaked with increasing $\alpha$. However, one should remember that the Schwarzschild mass has not much to do with the integrated energy density: it is defined by the value of the metric function $B(r)$ at the edge of the star $r=r_f$. The behavior of $B(r)$ is also presented in Fig.~(\ref{fig:neutron_prof1}) and~(\ref{fig:neutron_prof2}). In the latter figure, the $B(r)$ curve seems to be acting rather strangely in the case $\alpha = 10^{-19}$ m$^2$: the mass parameter becomes temporarily negative inside the star. This is a general feature that starts to take place at the higher densities the smaller $\alpha$ is, but never in the case of general relativity. Before the edge of the star the mass parameter becomes positive, and the Schwarzschild mass is never negative. 

Naturally, at some high enough densities we expect to encounter the situation where $2m(r)/r \ge 1$ $-$ the black hole. In general relativity, a neutron star with mass exceeding the TOV limit is predicted to collapse due to the gravitational instability to these densities. In the $\alpha$ models, however, the collapse seems to be disrupted by a stable region like in the case of white dwarfs. Thus to form a black hole, the neutron star needs to absorb a considerable amount of mass from a partner star. This result, nonetheless, is questionable: the equation of state $-$ the Fermi equation of state of degenerate neutrons alone, throughout the star $-$ is expected to be far from reality.

%

%
%

\section{Conclusions}
\label{sec:summary}

In this paper we have studied the interior spacetimes of white dwarfs and neutron stars in Palatini $f(R)$ gravity theories. A change in the structure of compact stars when compared to general relativity is expected both in the case of increasing \emph{and} decreasing function $F \equiv \partial f / \partial R$: in the former case, this is due to the high densities (and thus large $R$) inside compact objects, and in the latter case due to the fact that the equation of state of the matter approaches the equation of state of radiation (and thus yields small $R$). 

To make a first sketch, we have used the equation of state of a single species spin-$\half$ Fermi gas from the center to the edge of the star, which is rather a rude simplification and could be taken to the next level to account more realistically for the structure of white dwarfs and neutron stars. After defining the equation of state, we are able to solve numerically the field equations of the particular $f(R)$ model and thus track the density/pressure and spacetime curvature profiles of the stars. In all Palatini models, the interior solution is smoothly connected to the exterior Schwarzschild - de Sitter solution. Our results show how starting from a certain central density leads to different Schwarzschild masses in different $f(R)$ models. In this paper we have mainly concentrated on the model $f(R) = R + \alpha R^2$ and shown how 1) the Chandrasekhar limit disappears if $\alpha \gtrsim 10^{-16}$ m$^2$ and 2) the higher the parameter $\alpha$ is, the lower Schwarzschild mass and central density the Tolman-Oppenheimer-Volkoff limit is shifted to, and it seems to disappear for $\alpha \gtrsim 10^{-19}$ m$^2$. In the case of the model $f(R) = R - \mu^4/R$ we found out that there is no difference to the predictions of general relativity even at extremely high densities, unless the parameter $\mu^4 >> 1$ m$^{-4}$, that is unnaturally large.

The stability curves of white dwarfs and neutron stars can be easily produced in different $f(R)$ models, and we will go through a selection of models \emph{and} use improved equations of state in an upcoming article. The results can be used to restrict the parameter space of $f(R)$ models due to observations on binary systems and supernova explosions. While this paper was being completed, two papers,~\cite{Cooney:2009rr} and~\cite{Babichev:2009fi}, on relativistic stars in metric $f(R)$ theory also appeared on the arXiv.

%
%

\begin{acknowledgments}
We thank Kari Enqvist and Kimmo Kainulainen for the most valuable comments and discussions. This work was partially supported by a grant from the Magnus Ehrnrooth Foundation and the Waldemar von Frenckell Foundation. Finally, we also acknowledge the Marie Curie Research Training Network HPRN-CT-2006-035863.
\end{acknowledgments}

%
%

\end{document}